\documentstyle[aps,twocolumn]{revtex}
\begin{document}
\draft
\title{Disordered systems and the metal-insulator Transition: A super universality
class.}
\author{E. Hofstetter}
\address{Blackett Laboratory, Imperial College, London SW7 2BZ, United Kingdom}
\author{(To be published in Phys. Rev. B)}
\author{\small\parbox{14.5cm}{\vspace{0.5cm} \small
The critical behaviour of three-dimensional disordered systems is investigated 
by analysing the spectral fluctuations of the energy spectrum. Our results 
suggest that the initial symmetries (orthogonal, unitary and symplectic) are 
broken by the disorder {\it at} the critical point. The critical behaviour, 
determinedby the symmetry {\it at} the critical point, should therefore be 
independent of the previous invariances and be described by a ``super'' 
universality class. This result is strongly supported by the fact that we obtain 
the same critical exponent $\nu \simeq 1.35$ in the three cases: orthogonal, 
unitary and symplectic.
\vspace{0.5cm}
}}{}
\maketitle
\narrowtext
The transport properties in three-dimensional disordered systems are a very rich
topic that already has attracted considerable research \cite{MKK}. In particular
the system exhibits a metal-insulator transition (MIT) as a function of the 
disorder. Although it seems clear now that the MIT is a second-order phase
transition many points still remain unclear. One of them is the influence
of the symmetry on the MIT. It is generally assumed \cite{MKK} that the critical
behaviour at the MIT can be cast into three different universality classes 
according to some general symmetries of the system: orthogonal (with time 
reversal symmetry, O(N)), unitary (without time reversal symmetry, e.g. with a 
magnetic field, U(N)) and symplectic (with spin orbit-coupling, Sp(N)). One then 
expects different critical exponents related to the MIT for the three different 
universality classes. Surprisingly, in spite of the change of universality 
class, the same value of the critical exponent has been found, numerically, 
with and without a magnetic field \cite{Kra1,Kra2}. The MIT being a transition
between a chaotic (metallic regime) and a non-chaotic system (insulating 
regime) \cite{DG,BA}, a convenient way to study this problem is to resort to 
random matrix theory (RMT) and energy level-statistics (ELS). In RMT the 
statistics of the energy spectrum are generally described by three different
ensembles, Gaussian orthogonal (GOE), unitary (GUE) and symplectic (GSE) 
ensembles depending upon the symmetries mentioned above. It has been shown 
\cite{Eh1,Shkl}, in the case without magnetic field, that besides the two 
expected statistics, namely the GOE for the metallic regime and the Poisson 
ensemble (PE) for the insulating regime, a third statistic, called the critical
ensemble (CE), occurs only exactly {\it at} the critical point. The properties
of the CE will reflect, in particular, the shape of the critical states, which
are multifractals \cite{Mich}. It was recently proposed \cite{Eh2,Eh3} that 
a natural way to understand \cite{Kra1,Kra2} would be that the 
universality class of the system is determined by the symmetry of the system 
{\it at} the critical point with the previous results \cite{Kra1,Kra2} being 
interpreted as the sign of the independence of the universality class to the 
breaking of time-reversal symmetry {\it at} the critical point. We think
that when the system reaches the critical point, for increasing disorder, the 
${\rm O}(N)$ or ${\rm U}(N)$ symmetry is spontaneously broken. The reason 
for this break is obvious when one considers the localised regime. If the matrix
would be ${\rm O}(N)$  or ${\rm U}(N)$ invariant it would then be possible to 
construct delocalised states by linear combinations of the localised ones which 
is clearly not allowed. In RMT this 
phenomenon has led to the concept of a preferential basis. The distribution of 
the eigenvalues and the eigenvectors, which are independent for the GOE or GUE, 
become correlated when the disorder is increased and eventually break the 
${\rm O}(N)$ or ${\rm U}(N)$ invariance \cite{Pich}. Similar results concerning
the break of the ${\rm O}(N)$ invariance have been reported when studying
the two-level correlation function \cite{Aro}. Considering the spacing 
distribution $P(s)$ of the energy levels, previous results \cite{Eh3} suggest 
that the term inducing the eigenvalue-eigenvector correlation and that 
eventually breaks the symmetry appears, at the critical point, only in the large
$s$ behaviour of $P(s)$ and not in its small $s$ behaviour. The symmetry of the 
system is therefore reflected only in its large $s$ behaviour. However, in a 
recent paper \cite{ToS}, numerical evidence has been presented showing a small 
difference between the scaling properties of orthogonal and unitary systems
using more accurate raw data. But it has been shown \cite{Eh6} that an increase 
of the accuracy of the raw data with the transfer matrix method generates a 
{\it systematic} shift of the derived quantities and seems to give worse results,
casting doubts about the real accuracy of them. Moreover only one distribution
(box), for the site energy, was considered in \cite{ToS} whereas three different
(box, Gaussian and binary) were studied in \cite{Eh6}.
\par   
An interesting way to check this problem of symmetry is to study and to compare 
what happens at the critical point for the symplectic case or, in other words, 
to consider a system with spin-orbit coupling. It is generally assumed that the 
critical exponent for the symplectic case should be different ($\nu\simeq 1$) 
\cite{Ang} from the orthogonal or unitary case ($\nu\simeq 1.35$) although some 
recent numerical calculations \cite{To} have surprisingly given a critical 
exponent higher than expected in the symplectic case. In this paper we would 
like to show that the picture with the three universality classes to describe 
the MIT is probably inadequate and to propose a new way to understand the MIT, 
shedding new light on the problem.
\par
In order to investigate the MIT with spin-orbit coupling we consider the 
following spin-$\frac{1}{2}$ tight-binding Hamiltonian \cite{Eva}:
\begin{equation}
H=\sum_{j,\sigma} \varepsilon_{j,\sigma} |j,\sigma\rangle
\langle j,\sigma|+\sum_{j,j',\sigma,\sigma'}
V_{j,j'}^{\sigma,\sigma'} |j,\sigma\rangle
\langle j',\sigma'| \;,
\end{equation}
with the hopping matrix elements represented by
\begin{equation}
V_{j,j'}^{\sigma,\sigma'}= t^{0}{\bf 1} +i \mu \sum_{\gamma=1}^{3} t^{\gamma}
\mbox{\boldmath $\sigma$}_{\gamma} \;,
\end{equation}
where the sites $j$ are distributed regularly in three-dimensional (3D) space, 
e.g. on a simple 
cubic lattice. Only interactions with the nearest neighbours are considered. 
The site energy $\varepsilon_{j,\sigma}$ is described by a stochastic variable. 
In the present investigation we use a box distribution with variance 
$=\sqrt{W^{2}/12}$. $W$ represents the disorder and is the critical 
parameter. ${\bf 1}$ and $\mbox{\boldmath $\sigma$}_{\gamma}$ are the identity
and the Pauli matrices. We choose the coupling constant $\mu=1$, 
$t^{0}=1$ and $t^{\gamma}$ as independent random variables distributed 
uniformally between $[-\frac{1}{2},\frac{1}{2}]$ for each pair of lattice sites 
$j$ and $j'$. The Hermiticity and the time reversal symmetry impose 
$V_{j,j'}^{\sigma,\sigma'}=(V_{j',j}^{\sigma',\sigma})^{\ast}$ and 
$V_{j,j'}^{-\sigma,-\sigma'}=\sigma\sigma'(V_{j,j'}^{\sigma,\sigma'})^{\ast}$
respectively, with $\sigma,\sigma'=\pm 1$.
\par
Based on this Hamiltonian, the MIT in the presence of spin-orbit coupling is
studied by the ELS method, i.e. via the fluctuations of the energy spectrum
\cite{Eh1}. Starting from Eq. (1) the energy spectrum was computed by means of 
the Lanczos algorithm for systems of size $M \times M \times M$ with 
$M=13,\;15,\;17,\;19 \;{\rm and}\; 21$ and disorder $W$ ranging from 8 to 30.
The number of different realizations of the random site energies 
$\varepsilon_{j,\sigma}$ was chosen so that about $ 10^{5}$ eigenvalues were 
obtained for every pair of parameters $(M,W)$ for which only half of the 
spectrum around the band center is considered so that the results do not 
deteriorate due to the strongly localised states near the band edges. We
checked that we obtain the same results using bands of the energy spectrum ranging 
from 10 to 50\% (half spectrum). After unfolding the spectrum obtained, the 
fluctuations can be appropriately characterised by means of the spacing 
distribution $P(s)$ and the Dyson-Mehta statistics $\Delta_{3}(L)$. $P(s)$ 
measures the level repulsion, it is normalised, as is its first moment, because 
the spectrum is unfolded. $\Delta_{3}(L)$ measures the spectral rigidity. 
\par
Before studying what happens at the critical point we checked, for $P(s)$ and
$\Delta_{3}(L)$, that one finds, as expected, the GSE and the PE regimes for 
small and large disorder respectively. This can be seen in Fig. \ref{3} for 
$P(s)$ and Fig. \ref{5} for $\Delta_{3}(L)$.
\par
The next step is to find where the MIT takes place. For this, one uses 
that the quantities we are considering here, $P(s)$ and $\Delta_{3}(L)$, 
are scale invariant {\it at} the critical point. This is due to the fact that 
the MIT is a second order transition and that finite-size scaling laws apply 
close to the transition. So we calculate $\alpha_{r}(M,W)=\frac{1}{30} 
\int_{0}^{30} \Delta_{3}(L) dL$ as a function of $M$ and $W$, the critical 
disorder, $W_{c}$, being given by $W$ for which $\alpha_{r}(M,W)$ is independent 
of $M$ \cite{Eh4}. $\alpha_{r}(M,W)$ has been calculated with a step for $W$ of 
0.25 around the critical disorder and between 1 and 2 otherwise. But instead of 
using directly these data (raw data) we used the analytical properties of
$\alpha_{r}(M,W)$ for a finite system to fit a third order polynomial between 
$W=16$ and $W=26$ for each $M$. The results ($\alpha(M,W)$) are shown in 
Fig. \ref{1}. The raw data ($\alpha_{r}(M,W)$)
that where used for the fit are given in the inset of Fig. \ref{1}. This 
method is very interesting because, as we will see later, less computational 
effort is required and more accurate results are obtained. We can already see 
that the critical disorder
$W_{c}$ is accurately determined. One finds $W_{c}=21.75\pm 0.10$. This value
is higher than in the orthogonal, $W_{c}\simeq 16.5$ \cite{MKK}, and unitary 
case where it saturates, as a function of the strength of the magnetic field, 
around $W_{c}\simeq 18.75$ \cite{Eh3}. This is due to the weak localization, 
suppression of the weak localization, and antiweak localization phenomena, which
take place in the orthogonal, unitary, and symplectic cases
respectively \cite{Il}. It is then more difficult to localise the states, which 
means an upwards shift of $W_{c}$. With $W_{c}$ it is now possible to study 
$P(s)$ at the critical point. The first feature we would like to consider is 
the small $s$ behaviour of $P(s)$. In fact for numerical reasons what we 
calculate is the cumulative level-spacing distribution $I(s)=\int_{0}^{s} P(s') 
ds'$ which allows a better study of the small $s$ behaviour of $P(s)$. The 
results are plotted in Fig. \ref{2}. We have $P(s)\propto a s^{4}$ with 
spin-orbit coupling, $P(s)\propto b s^{2}$ with magnetic field and 
$P(s)\propto c s$ without. The behaviour is the same as in the metallic regime 
except for the prefactors $a$, $b$, and $c$, which are now higher, showing a 
decrease of the level repulsion due to the multifractal nature of the critical 
states \cite{Mich}. The different small $s$ behaviours obtained has been 
interpreted  
\cite{To,Isa} as a sign that the disorder does not modify the symmetry of the 
system and that we have, indeed, three different universality classes at the 
critical point. We claim here that this is not necessarily the case. Considering
a system with $N$ sites one sees \cite{Po} that the $s^{\beta}$ factor, 
with $\beta=1,2,4$, responsible for the small $s$ behaviour of $P(s)$ is, in 
fact, a geometrical factor related only to the size of the Hamiltonian, 
$N\times N$, $2N\times 2N$ and $4N\times 4N$ for orthogonal, unitary, and 
symplectic, respectively. In RMT, which corresponds to the metallic regime, this 
geometric factor certainly reflects the symmetry of the system but we do not 
believe it is necessarily the case for a system with large disorder. 
The distribution of the eigenvalues and the eigenvectors, which are independent 
for the GOE, GUE, or GSE, become correlated when the disorder is increased
and eventually break the ${\rm O}(N)$, ${\rm U}(N)$, or ${\rm Sp}(N)$ invariance. 
We think that the term inducing the eigenvalue-eigenvector correlation and 
which breaks the symmetry at the critical point appears first in the large $s$ 
behaviour of $P(s)$. For larger disorder, in the thermodynamic limit, this 
term will modify the small $s$ behaviour of $P(s)$ too, by cancelling the 
geometrical factor giving rise eventually to the PE. So following our 
argumentation what one should compare is the large $s$ behaviour of $P(s)$
for the three different ensembles at the critical point. The results are
reported in Fig. \ref{3}. As already shown \cite{Eh3} we see no 
difference between the orthogonal and the unitary case. The surprising
fact is that we obtain the same results for the symplectic case too (Fig. 
\ref{3}). According to what we wrote above this would mean that there is only 
one ``super'' universality class at the critical point in contradiction to the 
fact that one expects a different critical exponent for the symplectic case.
A way to check that is to calculate $\nu$ for the symplectic case. 
Starting with $\alpha(M,W)$ defined above, it was shown \cite{Eh4} that
$\alpha$ can be expressed as $\alpha(M,W)=f(M/\xi_{\infty}(W))$ with 
$\xi_{\infty}(W)=\xi|W-W_{c}|^{-\nu}$, the correlation length.
$\alpha(M,W)$, being analytical for a finite system, can be written around the
critical point as
\begin{equation}
\alpha (M,W) \simeq \alpha (M,W_{c})+ C |W-W_{c}| M^{1/\nu} \;.
\end{equation}
To perform the scaling procedure with the $\alpha_{r}(M,W)$ data, the range of 
$\alpha_{r}(M,W)$ values for various $M$ at any given disorder $W_{1}$ must 
overlap the range of $\alpha_{r}(M,W)$ values for various $M$ for at least one 
different disorder $W_{2}$. That would require us to compute $\alpha_{r}(M,W)$ for 
a very large number of $W$ values or for larger $M$. Using the method with the 
fit as 
described above this is not necessary because from the fit we derived all the
values we need for the scaling procedure. Using the values for $\alpha (M,W)$ 
given in Fig. \ref{1} we show in Fig. \ref{4} that $\alpha (M,W)$ can be, 
indeed, expressed by a scaling function $f(M/\xi_{\infty}(W))$. In the inset of 
Fig. \ref{4} the correlation length $\xi_{\infty}(W)$ is reported as a 
function of the disorder. We clearly see the divergence at $W_{c}\simeq 21.75$. 
The quality of the curves shows that using a fit to the raw data certainly
improves the results. This is also reflected when using Eq. (3) the value of
the critical exponent $\nu$ is derived. The tricky point is that the formula (3)
is valid only in the vicinity of the critical point but, on the other hand, close 
to the transition the numerical inaccuracies are largest. The task is therefore 
to find an adequate range for $|W-W_{c}|$ that satisfies these two constraints.
With the raw data the accuracy of the results as well as the value of $\nu$
is very sensitive to any change in the range of $|W-W_{c}|$. With the value of
$\alpha(M,W)$ from Fig. \ref{1} this is no longer the case. The quantities 
calculated are much more stable on a wider range of $|W-W_{c}|$ indicating a
better accuracy of our results. We obtain $\nu=1.36\pm 0.10$
for the critical exponent which in very good agreement with $\nu=1.35\pm 0.15$
for the orthogonal \cite{Eh4,Eh5} and $\nu=1.35\pm 0.20$ for the unitary
case \cite{Kra2}. These results seem strongly to support that the behaviour
of the MIT can be cast into one ``super'' universality class and not into three
as previously claimed. It has to be noted that recently, using a
generalized version of the Ando model \cite{An}, a value of $\nu=1.30\pm 0.20$
has been found for the symplectic case \cite{To} in good agreement with our 
results using the Evangelou-Ziman model \cite{Eva}. It is interesting to note
that the results at the critical point seems to be independent of the model
as already noted in two dimensions \cite{To1}.
\par
Concerning $\Delta_{3}(L)$ the results, Fig. \ref{5}, are more difficult to 
interpret. We know that the shape of $\Delta_{3}(L)$ contains a term linear in 
$L$ as well 
as a non-linear term $L^{\omega}$, with $\omega < 1$ \cite{Eh2}. As for 
$P(s)$, it seems we have two different behaviours. For the nonlinear term, one
obtains different curves for the ${\rm O}(N)$, ${\rm U}(N)$ and ${\rm Sp}(N)$
cases. This behaviour was already observed by Batsch {\it et al.} \cite{Isa}
in the case with a magnetic field. If now we consider the linear term, all the 
curves seem to fall onto one curve. It is interesting to note that the two-point
correlation function $R(s)$, from which $\Delta_{3}(L)$ can be obtained, has 
recently been calculated \cite{Aro}. It was shown that $R(s)$ is composed of 
two terms. One of them is related to the break of the ${\rm O}(N)$ invariance 
and gives rise to the linear term in $\Delta_{3}(L)$. So as for $P(s)$ it
seems that the break of the invariance and therefore the change of symmetry
is only reflected in one part of $\Delta_{3}(L)$, namely the linear term.
Numerically, even with an accuracy of $0.1$ on the critical disorder, it is
quite difficult to see what happens for large $L$ and more work clearly
needs to be done to increase the accuracy of the results.
\par
In conclusion we think that the picture, with the three universality classes, 
${\rm O}(N)$, ${\rm U}(N)$ and ${\rm Sp}(N)$, to describe the MIT is probably 
inadequate. This picture comes from a field-theoretical approach using a
standard $\sigma$ model \cite{Il,Ba}. But we think our results are not 
necessarily in contradiction with the field-theoretical approach. Indeed,
although these analytical results give indications as to the existence
of a MIT as well as information about the weak localized regime, they
say nothing about the critical regime (critical point, critical exponent).
The problem comes from the $2+\epsilon$ expansion used to solve the 
$\sigma$ model. This perturbative analysis is often unreliable for describing
the critical regime \cite{Cas} even near $D=2$. It is now well known that the 
$2+\epsilon$ expansion gives incorrect results for $D=3$ when applied to MIT
(Ref. \cite{Il}) and therefore should be used with great care. Moreover, it is 
far from obvious that this method is adapted for the case with magnetic field or 
spin-orbit coupling where the lower critical dimension is $<2$. In particular, 
the multifractal character of the critical states \cite{Mich}, which we think 
plays a crucial role in the description of the critical behaviour, is up to now 
completely beyond the scope of the $2+\epsilon$ expansion. In contrast,
interesting progresses have recently been made applying supersymmetry
techniques to the (non linear)$\sigma$ model to go beyond the perturbative 
analysis \cite{Fyo} although
the critical properties of the MIT still remain unresolved in the frame of the 
field theoretical approach \cite{Ba,Fyo}. Our results suggest that the large 
disorder breaks the ${\rm O}(N)$, ${\rm U}(N)$, or ${\rm Sp}(N)$ invariance at 
the critical point. But this break of symmetry is only reflected in some parts 
of $P(s)$ or $\Delta_{3}(L)$. Comparing these parts one finds that
the critical behaviour of the MIT is no longer described by three 
different universality classes but by one ``super'' universality class. 
This result is supported by the fact that one obtains the same critical 
exponent $\nu\simeq 1.35$ with and without magnetic field as well as with 
spin-orbit coupling. Moreover, Ohtsuki and Kawarabayashi recently showed 
\cite{Tom2} that the anomalous diffusion and the the fractal dimension $D(2)$ 
is the same for ${\rm O}(N)$, ${\rm U}(N)$, or ${\rm Sp}(N)$ at the MIT in 
agreement with our results. 
\\
Finally it is interesting to note that our results are in good agreement with, 
at least, some experiments, for the value of the critical exponent 
\cite{Loe,Hor} as well as the the absence of influence, at the MIT, of the 
magnetic field \cite{Hor,Krav} and the spin-orbit coupling \cite{Loe}. But
clearly further experiments need to be done to check these points carefully,
in particular to understand the different results obtained for uncompensated
and compensated semiconductors.
\par
\vspace{0.5cm}
{\bf Acknowlegment:} useful discussions with A. MacKinnon and T. Ohtsuki are
gratefully acknowledged.

\newpage
\begin{figure}
\caption{$\alpha$ as a function of $M$ and $W$. The critical disorder is given 
by the point for which $\alpha$ is independent of $M$. $W_{c}=21.75\pm 0.10$.
The raw data ($\alpha_{r})$ used for the fit to derive $\alpha$ are given in 
the inset}
\label{1}
\end{figure}
\begin{figure}
\caption{ln-ln plot of the cumulative level-spacing distribution $I(s)$ for 
small $s$. We have $P(s)\propto a s^{4}$ with spin-orbit coupling ($-\cdot-$), 
$P(s)\propto b s^{2}$ with magnetic field ($---$) and $P(s)\propto c s$ without
($\cdots$). The behaviour is the same as in the metallic regime (---) except 
for the prefactors $a$, $b$, and $c$, which are now higher.}
\label{2}
\end{figure}
\begin{figure}
\caption{Large $s$ behaviour of $P(s)$. $\ast$ and $+$ are the curves at $W=3$ 
and $W=100$, respectively, showing the metallic and the insulating regimes for
$P(s)$. At the critical point the curves are independent of the presence or 
absence of a magnetic field as well as of the spin-orbit coupling. In the inset
there is a ln plot of the tail of $P(s)$.}
\label{3}
\end{figure}
\begin{figure}
\caption{In this figure we show that $\alpha (M,W)$ can be expressed by a 
scaling function $f(M/\xi_{\infty}(W))$. In the inset the correlation
length $\xi_{\infty}(W)$ is reported as a function of the disorder. We 
clearly see the divergence at $W_{c}\simeq 21.75$. }
\label{4}
\end{figure}
\begin{figure}
\caption{Dyson-Mehta statistics $\Delta_{3}(L)$. $\bullet$ is $\Delta_{3}(L)$
at $W=3$ and $W=100$. At the critical point one obtains three different sets of 
data which seem to merge into one linear set with increasing $L$.}
\label{5}
\end{figure}

\begin{references}
\bibitem{MKK}
B. Kramer and A. MacKinnon, Rep. Prog. Phys. {\bf 56}, 1469 (1993)
and references therein.
\bibitem{Kra1}
T. Ohtsuki, B. Kramer and Y. Ono, J. Phys. Soc. Jpn. {\bf 62}, 224 (1993).
\bibitem{Kra2}
M. Henneke, B. Kramer and T. Ohtsuki, Europhys. Lett. {\bf 27}, 389 (1994).
\bibitem{DG}
D. R. Grempel, R. E. Prange and S. Fishman, Phys. Rev. A {\bf 29}, 1639 (1984).
\bibitem{BA}
B. L. Alt'shuler, B. I. Shklovskii, Sov. Phys. JETP {\bf 64}, 127 (1986);
\bibitem{Eh1}
E. Hofstetter and M. Schreiber, Phys. Rev. B {\bf 48}, 16979 (1993).
\bibitem{Shkl}
B. I. Shklovskii, B. Shapiro, B. R. Sears, P. Lambrianides and H. B. Shore,
Phys. Rev. B {\bf 47}, 11487 (1993).
\bibitem{Mich}
M. Schreiber, Phys. Rev. B {\bf 31}, 6146 (1985); M. Schreiber and 
H. Gru\ss bach, Phys. Rev. Lett. {\bf 67}, 607 (1991).
\bibitem{Eh2}
E. Hofstetter and M. Schreiber, Phys. Rev. Lett. {\bf 73}, 3137 (1994).
\bibitem{Eh3}
E. Hofstetter, Phys. Rev. B {\bf 54}, 4552 (1996).
\bibitem{Pich}
J. -L. Pichard and B. Shapiro, J. Phys. {\bf 4}, 623 (1994).
\bibitem{Aro}
A. G. Aronov adn A. D. Mirlin, Phys. Rev. B {\bf 51}, 6135 (1995);
C. M. Canali and V. E. Kravtsov, Phys. Rev. E {\bf 51}, 5185 (1995).
\bibitem{ToS}
K. Slevin and T. Ohtsuki, Phys. Rev. Lett. {\bf 78}, 4083 (1997).
\bibitem{Eh6}
E. Hofstetter, PhD Thesis, Universit\"{a}t Mainz, 1994.
\bibitem{Ang}
A. MacKinnon, private communication.
\bibitem{To}
T. Kawarabayashi, T. Ohtsuki, K. Slevin and Y. Ono, Phys. Rev. Lett.
{\bf 77}, 3593 (1996).
\bibitem{Eva}
S. N. Evangelou and T. A. Ziman, J. Phys. C {\bf 20}, L235 (1987).
\bibitem{Eh4}
E. Hofstetter and M. Schreiber, Phys. Rev. B {\bf 49}, 14726 (1994).
\bibitem{Il}
I. V. Lerner, in {\it Localisation 1990}, eds. K. A. Benedict,
J. T. Chalker, Institute of Physics Conf. Ser. {\bf 108}, 53 (1991)
and references therein.
\bibitem{Isa}
M. Batsch, L. Schweitzer, I. Kh. Zharekeshev and B. Kramer,
Phys. Rev. Lett. {\bf 77}, 1552 (1996).
\bibitem{Po}
C. E. Porter, ed., {\it Statistical Theories of Spectra: Fluctuations}
(Academic, New-York 1965).
\bibitem{Eh5}
E. Hofstetter and M. Schreiber, Europhys. Lett. {\bf 21}, 933 (1993).
\bibitem{An}
T. Ando, Phys. Rev. B {\bf 40}, 5325 (1989).
\bibitem{To1}
T. Ohtsuki and Y. Ono, J. Phys. Soc. Jpn. {\bf 64}, 4088 (1995).
\bibitem{Ba}
B. L. Altshuler and B. D. Simons in: {\it Mesoscopic Quantum Physics}; 
Proceedings of Les Houches, Session LXI, eds. E. Akkermans, G. Montambaux and
J. -L. Pichard (North Holland, Amsterdam, 1996) p.1.
\bibitem{Cas}
G. E. Castilla and S. Chakravarty, Nucl. Phys. B {\bf 485}, 613 (1997)
and references therein.
\bibitem{Fyo}
Y. V. Fyodorov in: {\it Mesoscopic Quantum Physics}; 
Proceedings of Les Houches, Session LXI, eds. E. Akkermans, G. Montambaux and
J. -L. Pichard (North Holland, Amsterdam, 1996) p.493 and references therein. 
\bibitem{Tom2}
T. Ohtsuki and T. Kawarabayashi, J. Phys. Soc. Jpn. {\bf 66}, 314 (1997).
\bibitem{Loe}
H. Stupp, M. Hornung, M. Lakner, O. Madel and H. v. L\"{o}hneysen,
Phys. Rev. Lett. {\bf 71}, 2634 (1993).
\bibitem{Hor}
M. Hornung, A. Ruzzu, H. G. Schlager, H. Stupp and H. v. L\"{o}hneysen, 
Europhys. Lett. {\bf 28}, 43 (1994).
\bibitem{Krav}
S. V. Kravchenko, W. E. Mason, G. E. Bowker, J. E. Furneaux, V. M. Pudalov
and M. D'Iorio, Phys. Rev. B {\bf 51 } 7038 (1995).
\end{references}
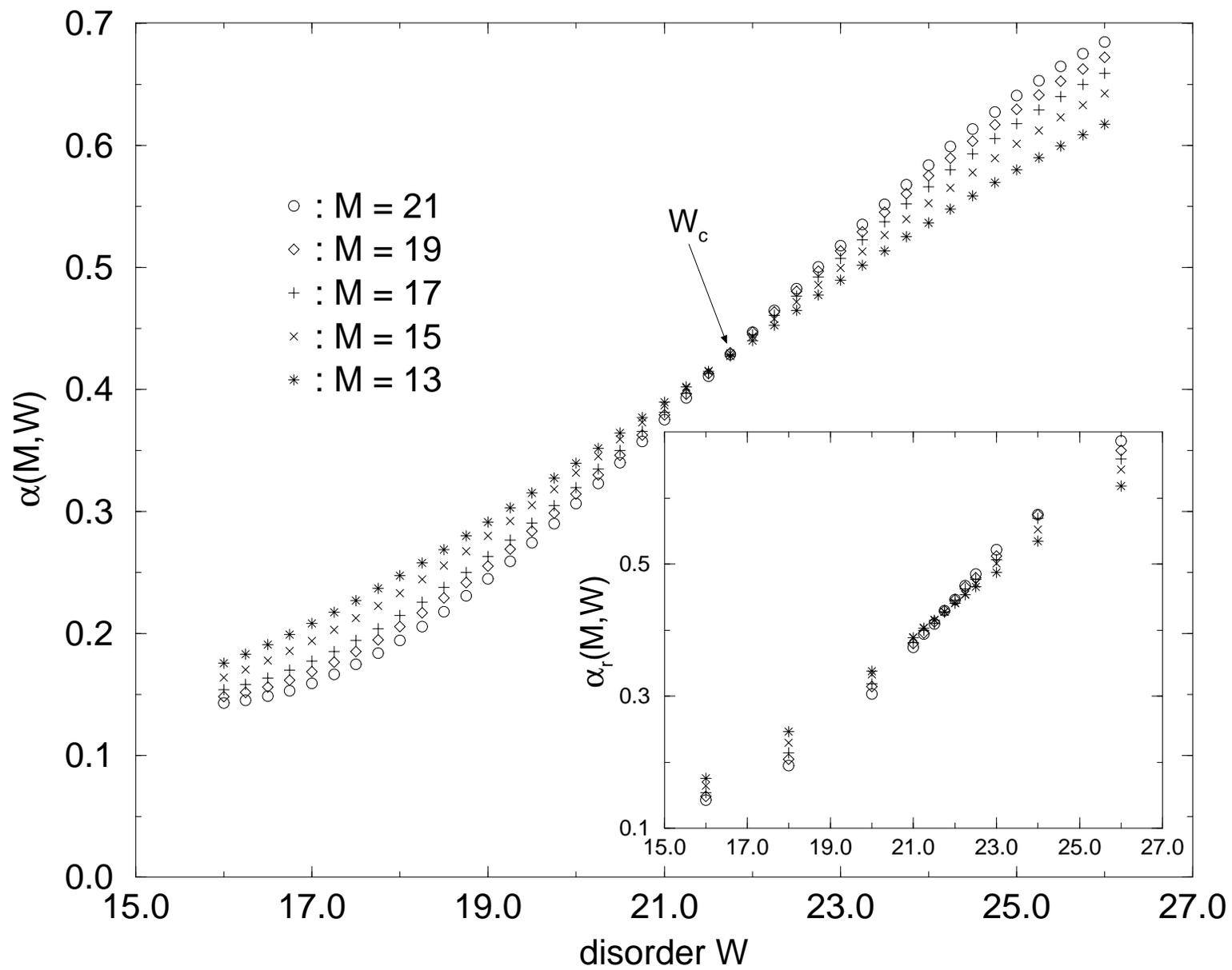
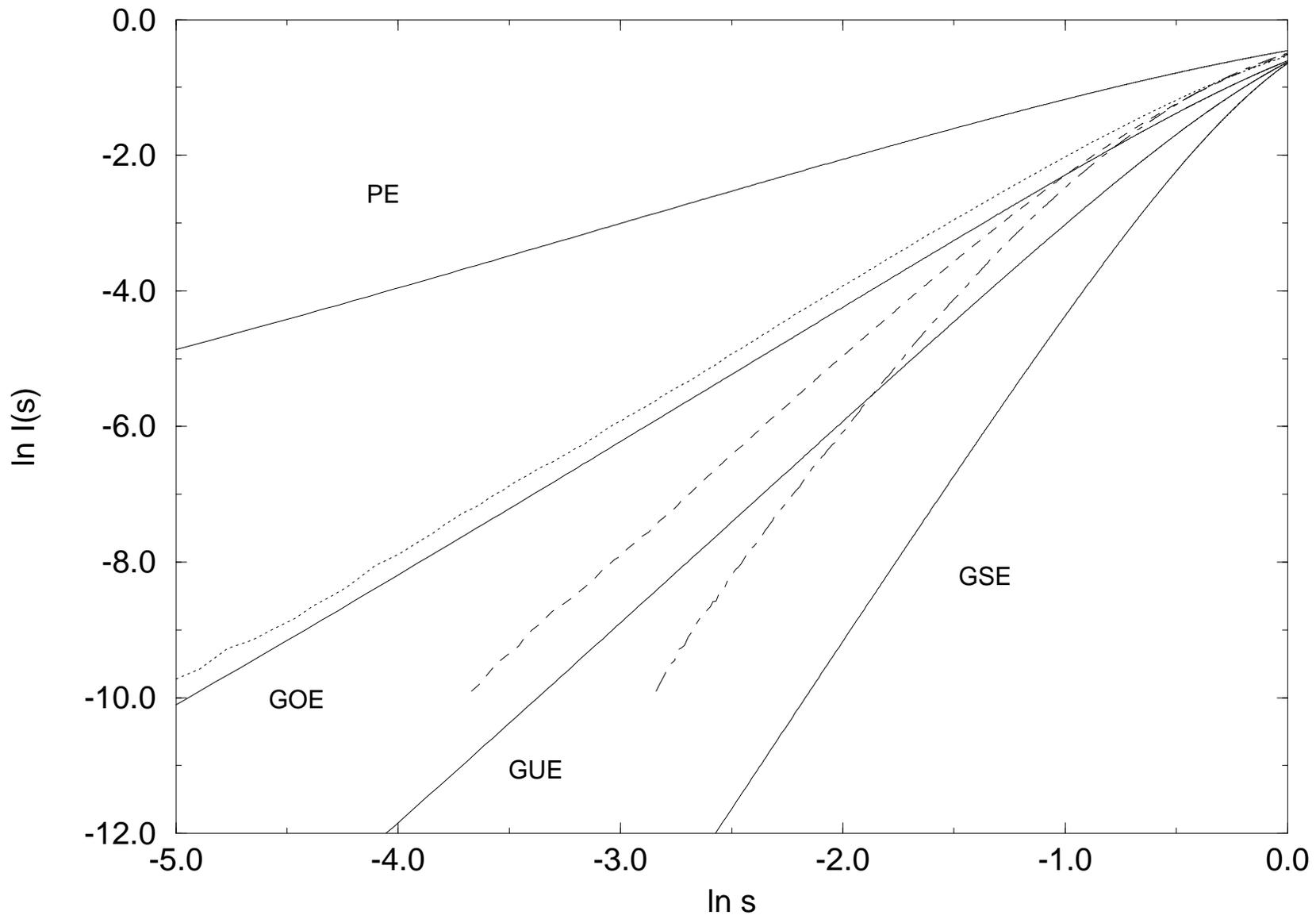
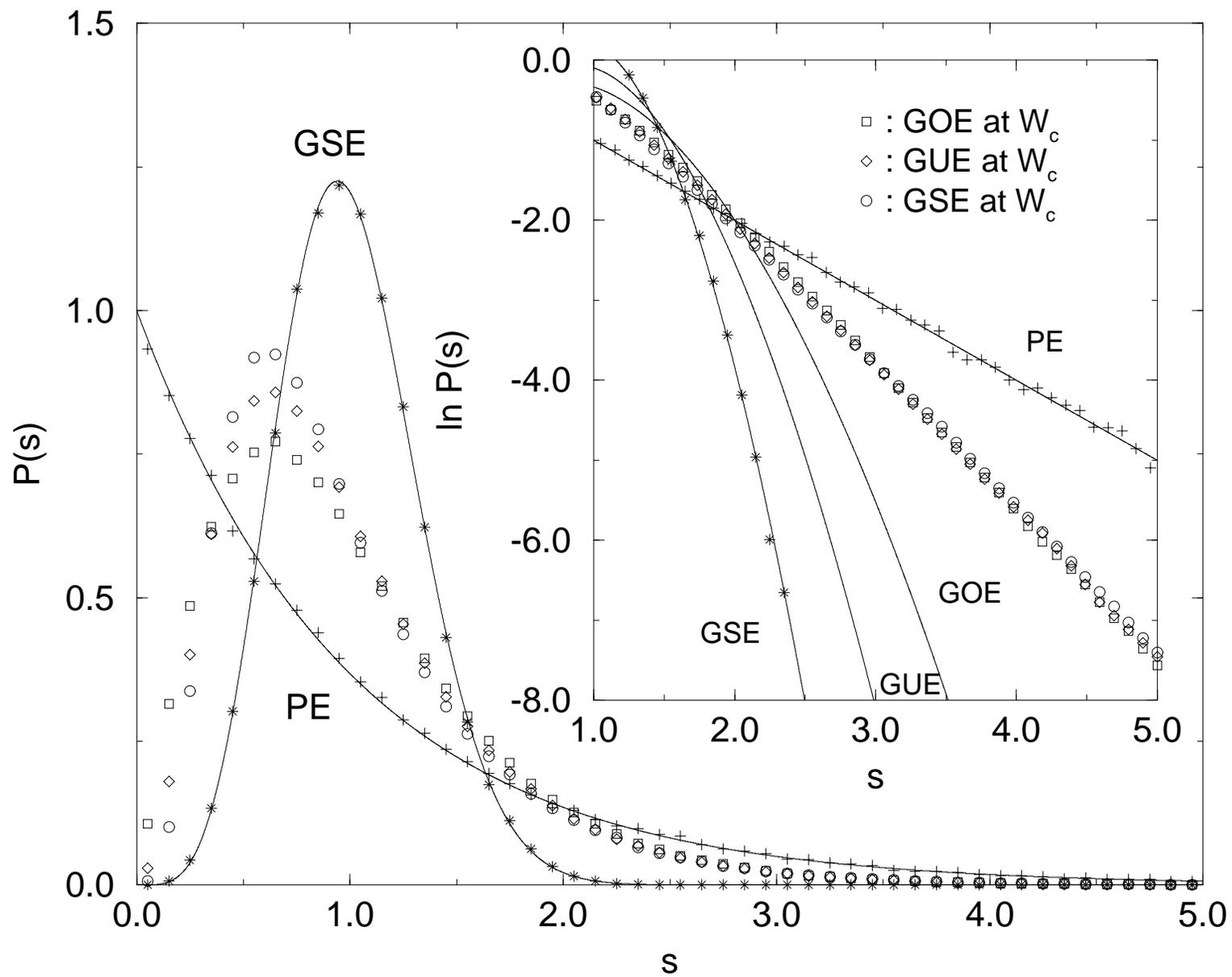
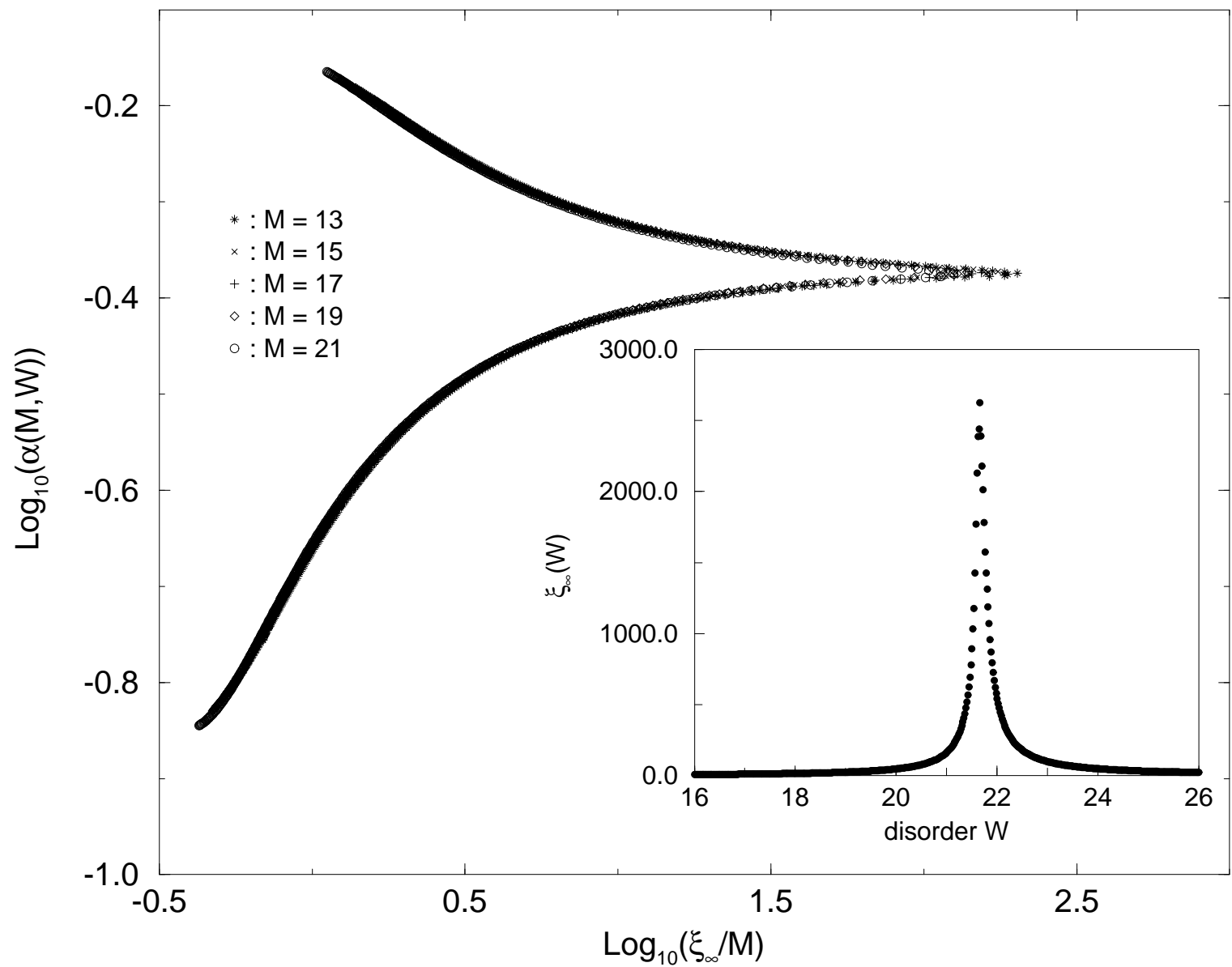
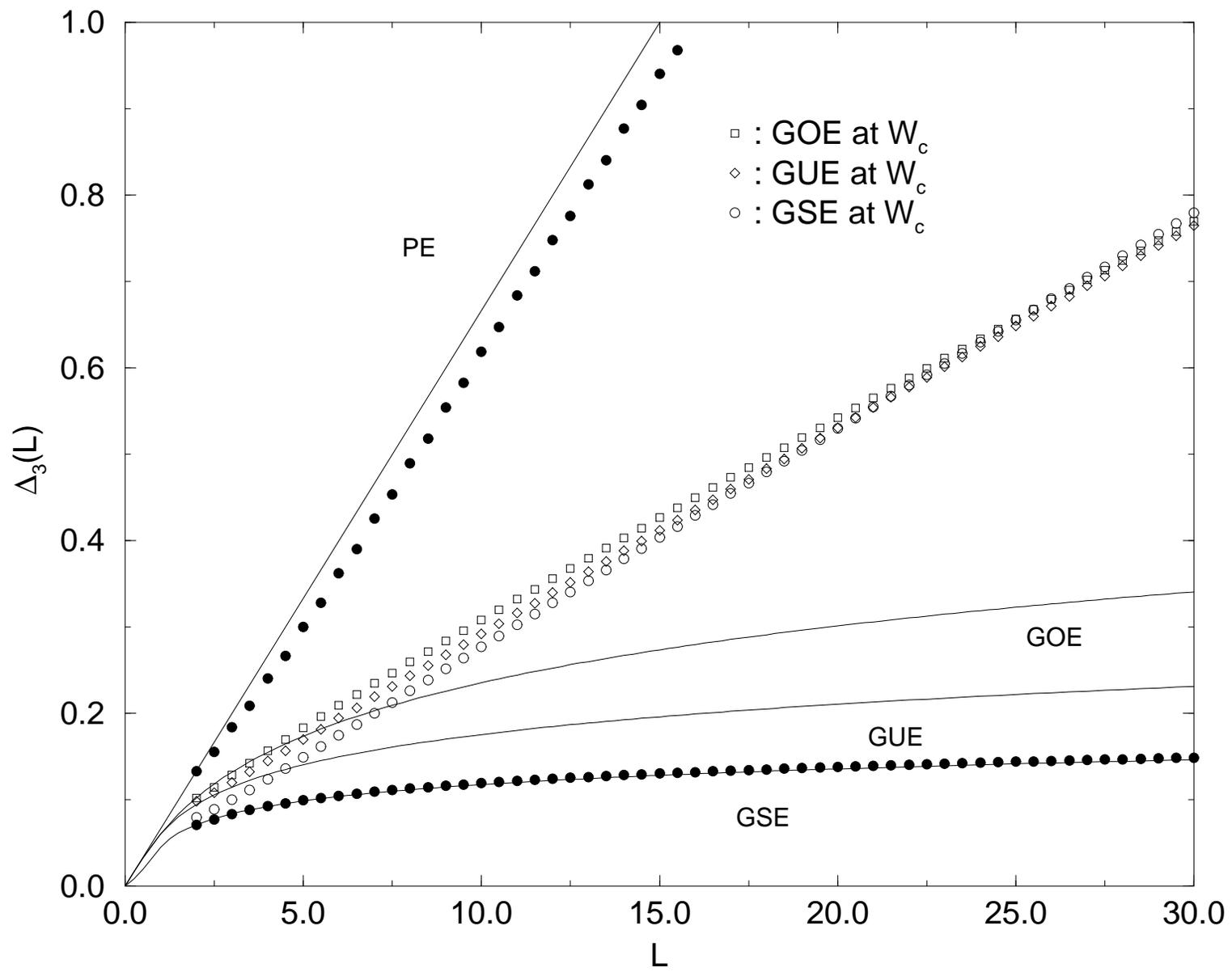
\end{document}